\documentstyle[twoside,fleqn,espcrc2]{article}

\newcommand{\AmS}{{\protect\the\textfont2
  A\kern-.1667em\lower.5ex\hbox{M}\kern-.125emS}}

\makeatletter

\def\lsim{\compoundrel<\over\sim}
\def\compoundrel#1\over#2{\mathpalette\compoundreL{{#1}\over{#2}}}
\def\compoundreL#1#2{\compoundREL#1#2}
\def\compoundREL#1#2\over#3{\mathrel{\vcenter{\hbox
{$\m@th\buildrel{#1#2}\over{#1#3}$}}}}
\makeatother
\title{New Predictions for 
Neutrino Telescope Event Rates}
\author{Raj Gandhi\address{Mehta Research Institute, 
10, Kasturba Gandhi Marg,
Allahabad 211002, India}\llap, 
Chris Quigg\address{Fermi National 
Accelerator Laboratory,
Batavia, IL 60510 USA}\llap, M. H. Reno\address{Department 
of Physics and Astronomy, University of Iowa,
Iowa City, IA 52242 USA}\llap, 
and Ina 
Sarcevic\address{Department of 
Physics, University of Arizona,
Tucson, AZ 85721 USA}\thanks{Talk presented by I. Sarcevic.}}
\begin{document}
\begin{abstract}
Recent measurements of the small-$x$ deep-inelastic regime at HERA
translate to new expectations for the neutrino-nucleon cross section
at ultrahigh energies. We present event rates 
for large underground neutrino telescopes
based on the new cross
section for a variety of models of neutrino 
production in Active Galactic
Nuclei, and we compare these rates with earlier 
cross section calculations.
\end{abstract}

\maketitle

Neutrino telescopes such as AMANDA, BAI\-KAL, DUMAND II and 
NESTOR \cite{Detect} have
the potential to extend the particle physics frontier beyond the
standard model, as well as probe stars and galaxies.
At ultrahigh energies ($E_\nu > 1$ TeV), neutrinos are decay products
of pions produced in cosmic ray interactions. 
Undeflected by magnetic fields and with long interaction lengths,
neutrinos can reveal information 
about astrophysical sources.  Gamma-rays, 
on the other hand, get absorbed by a few hundred gm of material.  
Active Galactic Nuclei (AGNs) \cite{review} may be prodigious sources
of high energy neutrinos as well as gamma-rays. 
Neutrino telescopes span a significant fraction of the sky at
all times, making the observation of neutrino interactions in or near
the detector feasible. If the most optimistic flux predictions are 
accurate, observations of AGNs via neutrino telescopes
may be imminent.

Here we present predictions of
event rates for several models of the AGN  
neutrino flux 
\cite{stecker,mannheim,sp}.  
We also compare the predicted rates with the atmospheric neutrino 
background (ATM), {\it i.e.}, 
neutrinos produced by cosmic ray interactions 
in the atmosphere \cite{volkova}.  
These rates reflect a new calculation of the
neutrino-nucleon cross section which follows from recent results from
the HERA $ep$ collider \cite{He}.  
To reduce the background from muons produced
in the atmosphere, it is sufficient to consider the 
upward-going muons produced below the detector via $\nu_\mu$
($\bar\nu_\mu$)-N interactions.  
We also give predictions for downward-moving 
(contained) muon event rates 
due to $\bar\nu_ee$ interactions in the
PeV range.

The importance of the HERA experimental results for neutrino-nucleon
cross sections is in measurements of structure functions at small
parton momentum fractions $x$ and large momentum transfers $Q$. At
ultrahigh energies, the 
$\nu N$ cross section is dominated by $Q\sim M_W$,
the mass of the $W$-boson. Consequently, 
$x\sim M_W^2/(2ME_\nu\langle y \rangle)$, 
in terms of 
nucleon mass $M$, 
and $x$
decreases as the incident neutrino energy $E_\nu$ increases.
HERA results cover the interval $10^{-4}\lsim x\lsim 10^{-2}$ and
8.5 GeV$^2 \lsim Q^2\lsim 15 $ GeV$^2$ \cite{He}, and guide theoretical
small-$x$ extrapolations of the parton distributions at even smaller
values of $x$. Compared to pre-HERA cross section calculations \cite{Rq},
the new cross section calculation is
approximately a factor of four to ten times larger 
at the highest energies ($E_\nu=$10$^9$ TeV) \cite{Gqrs}.
The range of values reflects different parton distribution function 
parameterizations and extrapolations below $x=10^{-5}$, all consistent
with the data at higher $x$.
Since the larger cross section also implies greater attenuation of 
neutrinos in the Earth, 
upward-muon event
rates for neutrino energy thresholds in 
the 1-10 TeV range are only $15\%$ larger than 
previous results based on old
cross sections \cite{Rq}.

The attenuation of neutrinos in the Earth 
is described by a shadow
factor $S(E_\nu)$, equivalent to the
effective solid angle for upward muons, normalized to $2\pi$:
\begin{equation}
{d S(E_\nu)\over d\Omega}={1\over 2\pi}
\exp \Bigl( -z(\theta )N_A \sigma_{\nu N}(E_\nu) \Bigr),
\end{equation}
where $N_A=6.022\times 10^{23}$ mol$^{-1}=6.022\times 10^{23}$ cm$^{-3}$
(water equivalent) is Avogadro's number, and $z(\theta)$ is the column
depth of the earth, in water equivalent units, which depends on zenith 
angle \cite{prem}.
The probability 
that the neutrino
converts to a muon that arrives at the detector
with $E_\mu$ larger than the threshold energy $E_\mu^{\rm min}$ is 
proportional to the cross section: 
\begin{equation}
P_\mu(E_\nu,E_\mu^{\rm min}) = \sigma_{\rm CC}(E_\nu) N_A \langle
R(E_\nu,E_\mu^{\rm min} )\rangle ,
\end{equation} 
where the average muon range in rock is 
$\langle R\rangle$ \cite{slip}. A more detailed discussion appears in
Ref. \cite{Gqrs}.
 
The diffuse flux of AGN neutrinos, 
summed over all AGN sources, is isotropic, so
the event rate is
\begin{eqnarray}
{\rm Rate} = A 
\int dE_\nu P_\mu(E_\nu,E_\mu^{\rm min}) 
S(E_\nu){dN_\nu\over dE_\nu} ,
\end{eqnarray}
given a neutrino spectrum $dN_\nu/dE_\nu$ and detector cross sectional
area $A$. As the cross section increases, $P_\mu$ increases, but the
effective solid angle decreases.

In Tables 1  and 2, we show the event rates for a detector with $A=0.1$
km$^2$ for $E_\mu^{\rm min}=1$ TeV and $E_\mu^{\rm min}=10$ TeV,
respectively. These event rates are for
upward muons and antimuons with two choices of parton distribution
functions: EHLQ parton distribution functions \cite{Ehlq} used in
Ref. \cite{Rq}, and the parton distributions parameterized by
the CTEQ Collaboration \cite{CTEQ}, coming from a global fit 
that includes 
the 
HERA data. The muon range is that of Ref. \cite{slip}.

\begin{table}
\caption{
Number of upward $\mu+\bar{\mu}$
per year per steradian for $A=0.1$ km$^2$ and $E_\mu^{\rm min}= 1$ TeV.}
\begin{center}
\begin{tabular}{||l|c|c||}
\hline
Fluxes & EHLQ & CTEQ-DIS \\ \hline
AGN-SS \cite{stecker} & 82 & 92  \\ \hline
AGN-NMB \cite{mannheim} & 100 & 111  \\ \hline
AGN-SP \cite{sp} & 2660 & 2960 \\ \hline
ATM \cite{volkova}& 126 & 141 \\ \hline
\end{tabular}
\end{center}
\end{table}

\begin{table}
\caption{As in Table 1, but for 
$E_\mu^{\rm min}=10$ TeV.}
\begin{center}
\begin{tabular}{||l|c|c||}
 \hline
Fluxes & EHLQ & CTEQ-DIS \\ \hline
AGN-SS \cite{stecker} & 46  & 51  \\ \hline
AGN-NMB \cite{mannheim} & 31 & 34  \\ \hline
AGN-SP \cite{sp} & 760 & 843 \\ \hline
ATM \cite{volkova}& 3 & 3  \\ \hline
\end{tabular}
\end{center}
\end{table}

The representative fluxes in Tables 1 and 2 can be approximated by a
simple power law behavior for $E_\nu<100$ TeV. For $dN/dE_\nu\propto
E^{-\gamma}$, the fluxes can be approximated by $\gamma=0$ (AGN-SS),
$\gamma = 2$ (AGN-NMB and AGN-SP) and $\gamma = 3.6$ (ATM).
The AGN-SP rate is large compared to the AGN-NMB rate because additional 
mechanisms are included.  
Flux limits from the Fr\'ejus experiment
are inconsistent with the SP flux for 1 TeV$< E_\nu <$ 10 TeV 
\cite{frejus}.

The flatter neutrino spectra
have larger contributions to the event rate
for muon energies away from the
threshold muon energy than the steep atmospheric flux.
For the 10 TeV muon energy threshold, the atmospheric neutrino
background is significantly reduced.

Finally we consider event rates from electron neutrino and antineutrino
interactions. For $\nu_eN$ (and $\bar{\nu}_eN$) interactions, the
cross sections are identical to the muon neutrino (antineutrino) nucleon
cross sections. 
Because of the rapid energy loss or 
annihilation of electrons and positrons,
it is generally true that only contained events can be observed.  
Since electron neutrino fluxes are small, 
unrealistically large effective volume
is needed to get measurable event rates.  
The exception is at $E_\nu=6.3$ PeV,
precisely the energy for resonant $W$-boson production in $\bar{\nu}_e
e$ collisions.
The contained event rate for resonant $W$ production is
\begin{equation}
{\rm Rate} = {10\over 18} V_{\rm eff}  N_A
\int dE_{\bar{\nu}} \sigma_{\nu e}(E_\nu) 
S(E_{\bar{\nu}}){dN\over dE_{\bar{\nu}}} .
\end{equation}
We show event rates for resonant $W$-boson production in Table 3.

\begin{table}
\caption{
Downward resonance $\bar\nu_e e\rightarrow W^-$ events per 
year per steradian for a detector with effective volume $V_{\rm eff}=1$ 
km$^3$ 
together with the potential downward (upward) background from 
$\nu_\mu$ and $\bar\nu_\mu$ interactions above 3 PeV.}
\begin{center}
\begin{tabular}{||l|c|c||}
\hline
Mode & AGN-SS \cite{stecker} & AGN-SP \cite{sp} \\ \hline
$W\rightarrow \bar{\nu}_\mu \mu$ & 6 & 3 \\ \hline
$W\rightarrow {\rm hadrons}$ & 41 & 19 \\ \hline
$(\nu_\mu,\bar\nu_\mu)$ N CC  & 33 (7) & 19 (4) \\ \hline
$(\nu_\mu,\bar\nu_\mu)$ N NC  & 13 (3) & 7 (1) \\ \hline
\end{tabular}
\end{center}
\end{table}

From Table 3 we note that a 1 
km$^3$ detector with energy threshold in PeV range 
would be suitable for 
detecting resonant $\bar\nu_e \rightarrow W$ events,
however, the $\nu_\mu N$ background may be difficult to overcome.
By placing the detector a few km underground, one can reduce 
atmospheric-muon 
background, which is 5 events per year per steradian at the surface of 
the Earth.  

In summary, we find that detectors such as 
DUMAND II, AMANDA, BAIKAL 
and NES\-TOR have a very good chance 
of being able to test different models for
neutrino production in the AGNs \cite{review}.    
For $E_{\mu}^{\rm min}=1$ TeV, we find that 
the range of theoretical fluxes lead to event rates of 
900-29,600
upward-moving 
muons/yr/km$^2$/sr
originating from 
the diffuse AGN neutrinos, with the atmospheric background of 1400 
events/yr
/km$^2$/sr. 
For $E_{\mu}^{\rm min}=10$ TeV, 
signal to background ratio becomes even better, 
with signals being on the order of 500-8,400 events/yr/km$^2$/sr,
a factor $\sim$20-300 higher than the background rate.  
For neutrino energies above $3$ PeV there is significant contribution 
to the muon rate due to the $\bar \nu_e$ interaction with electrons, 
due to the W-resonance contribution.  We find that 
acoustic detectors with 3 PeV threshold and with 
effective volume of 
0.2 km$^3$, such as DUMAND, would detect 
48 hadronic 
cascades per year 
from 
$W \rightarrow$ hadrons, 7 events from 
$W \rightarrow \mu \bar\nu_{\mu}$ and 
36 events from $\nu_{\mu}$ and 
$\bar \nu_{\mu}$ interactions with virtually no background from 
ATM neutrinos.  

\end{document}